\documentclass{article}
\usepackage{spconf,amsmath,graphicx, hyperref,xcolor,colortbl,subfig, comment}
\hypersetup{
    colorlinks=true,
    linkcolor=blue,
    filecolor=magenta,      
    urlcolor=cyan,
}

\title{DNSMOS: A Non-Intrusive Perceptual Objective Speech Quality metric to evaluate Noise Suppressors}
\name{Chandan K A Reddy, Vishak Gopal, Ross Cutler}
\address{Microsoft Corporation, Redmond, WA\\
    chkarada@microsoft.com, vishak.gopal@microsoft.com, rcutler@microsoft.com}
\begin{document}
%
\maketitle
\begin{abstract}
Human subjective evaluation is the ``gold standard'' to evaluate speech quality optimized for human perception. Perceptual objective metrics serve as a proxy for subjective scores. The conventional and widely used metrics require a reference clean speech signal, which is unavailable in real recordings. Previous no-reference approaches correlate poorly with human ratings and are not widely adopted in the research community. One of the biggest use cases of these perceptual objective metrics is to evaluate noise suppression algorithms. This paper introduces a multi-stage self-teaching based perceptual objective metric that is designed to evaluate noise suppressors. The proposed method generalizes well in challenging test conditions with a high correlation to human ratings.     
\end{abstract}
\section{Introduction}
\label{sec:intro}

Subjective evaluation of speech quality is the most reliable way to evaluate Speech Enhancement (SE) methods \cite{reddy2019scalable}. However, subjective tests are not scalable as they require a considerable number of listeners, the process is laborious, time-consuming, and expensive. Conventional objective speech quality metrics such as Perceptual Evaluation of Speech Quality (PESQ) \cite{941023}, Perceptual Objective Listening Quality Analysis (POLQA) \cite{polqa} and Signal to Distortion Ratio (SDR) are widely used to evaluate Speech Enhancement (SE) algorithms optimized for human perception. Some of these metrics are designed to predict the subjective Mean Opinion Score (MOS) obtained using the Telecommunication Standardization Sector of the International Telecommunication Union (ITU-T) Recommendation P.800 \cite{p800}. There are also the intrusive metrics requiring a reference clean speech to compute the quality score. However, they are shown to correlate poorly with human rating when used for SE tasks that involve perceptually invariant transformations \cite{reddy2019scalable}. Also, intrusive metrics cannot be used to evaluate real recordings when clean reference is unavailable in realistic scenarios.

ITU-T Recommendation P.563 is a non-intrusive technique and can directly operate on the degraded signal \cite{p563}. However, it was developed for narrow-band applications and works well on limited impairment types. Recently, Deep Neural Networks (DNN) based approaches have been proposed to estimate the speech quality scores \cite{dong2020attention, gamper2019intrusive, 8683175}. Some of these learning-based approaches use other objective metrics as the ground truth to train their speech quality predictor. A handful of published methods use MOS obtained using P.800 as the ground truth to train their models. In \cite{manocha2020differentiable}, the authors trained the model to identify the Just Noticeable Difference (JND). 
MOS predictors trained on actual human ratings are more reliable than the ones trained to predict other objective metrics like PESQ or POLQA. The accuracy and robustness of the learned models depend on the quality of the human labels and also the quantity and diversity of the audio clips. There is no large scale human-labeled data set publicly available to train a model that can robustly generalize to a variety of audio impairments, especially for SE task. Though the learning-based approaches have a higher correlation to human ratings when tested on a limited and small test set, none of these speech quality metrics are as widely used as some of the intrusive objective metrics. Most of the publications in the area of noise suppression and SE test their methods on synthetic noisy speech test sets using intrusive speech quality metrics such as PESQ, POLQA and SDR. 

In this paper, we have developed a robust objective perceptual speech quality metric called DNSMOS. It can be used to stack rank different Deep Noise Suppression (DNS) methods based on MOS estimates with great accuracy and hence the name DNSMOS. A simple Convolutional Neural Network (CNN) based model is trained using the ground truth human ratings obtained using ITU-T P.808 \cite{naderi2020open, reddy2020interspeech}. P.808 is an online subjective testing framework that is highly reproducible and shown to stack rank noise suppression models with high accuracy when each model is tested as an average over a statistically significant number of clips. However, P.808 ratings per clip can be noisy due to various factors. The absolute MOS per clip might vary for the same clip in different P.808 runs due to the inherent noise in human rating \cite{P1401}. The number of ratings per clip varies due to spam removal. Also, all the spammers might not be filtered due to the limitations of spam filtering logic in P.808 implementation. We train a multi-stage self-teaching model inspired by \textit{continual life long learning} \cite{parisi2019continual} to learn in the presence of noisy labels. We show that the multi-stage training significantly improves the correlation and generalizes remarkably well to other impairment types that are very different from what was used in the training set. We found the DNSMOS very useful in our audio/speech research as it generalizes well to a variety of speech impairments. Hence, we are providing the DNSMOS as an Azure service for other researchers to use. The details of the API are at \url{www.microsoft.com/en-us/research/dns-challenge/dnsmos}.

\section{Data and subjective ratings}
\label{sec:data}

\begin{figure}[t]
 \centering
  \subfloat[Distribution of MOS ratings]{\includegraphics[width=0.24\textwidth]{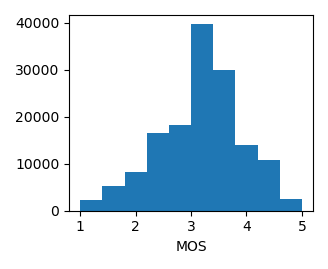}\label{fig:f1a}}
  \hfill
  \subfloat[Distribution of number of votes]{\includegraphics[width=0.24\textwidth]{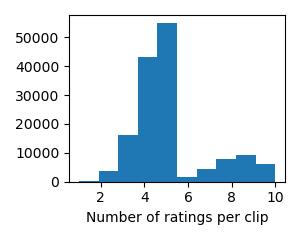}\label{fig:f1b}}
  \hfill
  \subfloat[Distribution of standard deviation per clip]{\includegraphics[width=0.25\textwidth]{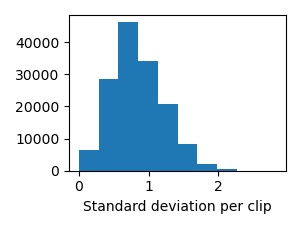}\label{fig:f1c}}
  
  \caption{Statistics of the training set}
\label{fig:res}
\end{figure}

We used 600 noisy speech test clips comprised of a combination of synthetic and real recordings. The synthetic test set is a combination of both non-reverberant and reverberant clips. The real recordings were captured in a variety of noise types and Signal to Noise Ratio (SNR) and target levels. The test set is comprised of over 100 noise types and speakers. More details about the creation of these test sets can be found in \cite{reddy2020interspeech}. 
These 600 clips were processed by over 200 different noise suppression methods. Some of these methods improved the overall speech quality, but they also introduced a variety of artifacts as a consequence of over or under suppression of noise, resulting in speech and noise distortions. The speech quality ratings of the processed clips varied from very poor (MOS=1) to excellent (MOS=5). The distribution of the MOS scores in the training data is shown in Figure \ref{fig:res}a. The scores are highly skewed with most ratings populated in the range 3\textless MOS\textless 4 and fewer ratings in both the tails. The distributions in Figure \ref{fig:res}a also shows that the current state of the art noise suppression techniques are not good enough to shift the overall distribution above MOS of 4. Figure \ref{fig:res}b shows the distribution of the number of ratings per clip in the training set. The skewness in the distribution is due to the fact that some of the ratings were removed in the process of removing spam ratings. Also, we used 10 ratings per clip in some of the experiments and 5 ratings per clip in others. Figure \ref{fig:res}c shows the distribution of standard deviation per clip. There are a significant number of clips with a standard deviation of over 1 MOS, which shows that the audio clips are highly subjective in nature. Increasing the number of ratings per clip will reduce the standard deviation, but also increases the cost. 

The subjective human ratings are obtained in several P.808 runs conducted over several months. Multiple noise suppression methods are compared in each P.808 run. Each P.808 run included the best performing noise suppressor, original noisy speech, and a couple of methods with intermediate perceptual quality from previous runs as anchors. Hence, some of the clips were rated multiple times. In total, we have about 120,000 audio clips with associated MOS scores as ground truth. The average length of each audio clip was about 9 seconds. In \cite{naderi2020open}, we show that P.808 is highly reproducible with high SRCC between runs when we average the ratings across clips processed by a given model. However, the absolute values of the scores vary between P.808 runs as the raters differ and humans tend to be inconsistent in perceptual tasks due to various biases. 

\section{DNSMOS}
\label{sec:model}

\subsection{Learning in the presence of label biases}
\label{ssec:biases}
The variation in the absolute values between P.808 runs might pose a challenge in training models that learn to predict MOS scores when the training data includes the same audio clip with different ratings. We know from the literature that DNN models can generalize well in spite of the noise for classification tasks \cite{reddy2019supervised,Rolnick2017DeepLI}. The variation in the absolute scores from different runs can be treated as noise in the labels. Also, the clips with fewer ratings can be treated as noise due to higher standard deviation. There are other unknown human biases that play a role in the variation of ratings. We initially took a direct approach where we trained a deep model by including all the available data without any modification. The model tends to learn the weighted average of the ratings across runs and also tries to average out other biases. This plain approach gave a good correlation with human ratings when tested on the data that was very similar to the training set. However, the correlation dropped when tested on a more challenging test set that was very different from the training data. 

We improved the generalization and accuracy of the predictor by training a few stages of self-teaching models. In \cite{kumarsequential}, a sequence of self-teachers were used to improve the generalizability of a classification task with binary cross-entropy loss in noisy and weakly labeled conditions. We use a similar approach for the regression problem at hand with Mean Squared Error (MSE) as the loss function. Unlike knowledge distillation where a larger teacher model is used to produce soft labels to train a smaller student model, we use the same model architecture for both teacher and the student, and hence the name self-teacher. The student model uses the weighted average of the predictions from one or multiple teacher models and the original human MOS ratings. 
The primary teacher model \(\textbf{\textit{M}}_0\) is trained until the loss saturates using the original ground truth human MOS ratings \(\textbf{\textit{r}}\). The student model \(\textbf{\textit{M}}_s\) at stage \textbf{\textit{s}}, is trained using the new target given by,
\begin{equation}
\hat{\textbf{\textit{r}}}_{s}=\alpha_0\textbf{\textit{r}}+\sum_{i=0}^{s-1}\alpha_{i+1}\hat{\textbf{\textit{r}}}_{i}
\end{equation}
where \(\hat{\textbf{\textit{r}}}_{i}\) is the prediction of model \(\textbf{\textit{M}}_i\) and \(\sum_{i=0}^{s}\alpha_{i}=1\). In this work, we restrict \textbf{\textit{s}} to 2. With some assumptions to the noise distribution, it can be theoretically proven that the \(\textbf{\textit{M}}_s\) is better than \(\textbf{\textit{M}}_0\) for \(\textit{s}>0\). Since the distribution of noise in our case is complex, we show through empirical experiments that the predicted MOS of \(\textbf{\textit{M}}_i\) is of higher accuracy than \(\textbf{\textit{M}}_0\) for appropriately chosen values of \(\alpha\).     
\subsection{Features}
\label{ssec:features}
Recently, researchers have seen success in learning features within the model for tasks such as SE \cite{alex2020real}, speech, and music synthesis \cite{tanaka2019wavecyclegan2} and to learn acoustic models \cite{7178847}. They show that using the time-domain waveform requires a larger model trained on a larger and diverse data set to ensure generalization. The ground truth MOS scores are obtained for audio clips with an average length of 9 secs sampled at 16 kHz. This leads to a very large input dimension if we are treating it as a vector and the model requires many layers to compress and extract input features. We used Log power Mel spectrogram as input feature as it correlates well with human perception and is proven to work very well for analyzing speech quality \cite{gamper2019intrusive}. For spectral features, we used a frame size of 20 ms with a hop length of 10 ms and 120 Mel frequency bands. The input features are then converted to dB scale. 

\subsection{Prediction model}
\label{ssec:models}
For predicting the MOS scores, we explored different configurations of CNN based models. The architecture for the best performing model is shown in Table \ref{tab:table1}, which is the architecture for all \(\textbf{\textit{M}}\). The input to the model is log power Mel Spectrogram with 120 Mel bands computed over a clip of length 9 secs sampled at 16 kHz with a frame size of 20 ms and hop length of 10 ms. This results in an input dimension of 900 x 120. The model was trained with a batch size of 32 using the Adam optimizer and MSE loss function until the loss saturated. We experimented by adding batch normalization layers after every Conv layer in Table 1. However, adding batch normalization reduces the prediction accuracy of low volume clips. Humans tend to give lower ratings to the clips with low amplitudes. We want the model to capture the variations in the target levels of the data. Hence, we avoid any kind of feature normalization. We also explored different network architectures including CNN followed by LSTM. The model in Table \ref{tab:table1} generalized the best and was of least complexity.    

\begin{table}[t!]
  \begin{center}
    \caption{DNSMOS Prediction Model}
    \label{tab:table1}
    \begin{tabular}{l|r}
      \textbf{Layer} & \textbf{Output dimension} \\
      \hline
      Input & 900 x 120 x 1\\
      \hline
      Conv: 32, (3 x 3), `ReLU' & 900 x 120 x 32 \\
      MaxPool: (2 x 2), Dropout(0.3) & 450 x 60 x 32 \\
      \hline
      Conv: 32, (3 x 3), `ReLU' & 450 x 60 x 32 \\
      MaxPool: (2 x 2), Dropout(0.3) & 225 x 30 x 32 \\
      \hline
      Conv: 32, (3 x 3), `ReLU' & 225 x 30 x 32 \\
      MaxPool: (2 x 2), Dropout(0.3) & 112 x 15 x 32 \\
      \hline
      Conv: 64, (3 x 3), `ReLU' & 112 x 15 x 64 \\
      GlobalMaxPool & 1 x 64 \\
      \hline
      Dense: 64, `ReLU' & 1 x 64 \\
      Dense: 64, `ReLU' & 1 x 64 \\
      Dense: 1  & 1 x 1 \\
      \hline
      
    \end{tabular}
  \end{center}
\end{table}

\begin{table}[!b]
  \begin{center}
    \caption{Correlation of DNSMOS with other widely used objective metrics}
    \label{tab:table2}
    \begin{tabular}{c|c|c|c|c}
      \textbf{  } & \textbf{PESQ} & \textbf{SDR} &\textbf{POLQA} &\textbf{DNSMOS (\(\textbf{\textit{M}}_0\))} \\
      \hline
      PCC &  0.78 & 0.23 & 0.79 & 0.93\\
      \hline
      SRCC &  0.82 & 0.25 & 0.84 & 0.94\\
      \hline
      
    \end{tabular}
  \end{center}
\end{table}

\section{Experimental Results}
\label{sec:training}
\subsection{Evaluation metric}
\label{ssec:evaluation}
Pearson Correlation Coefficient (PCC) or MSE between the predictions of the developed objective metric and the ground truth human ratings is commonly used to measure the accuracy of the model \cite{gamper2019intrusive, 8683175}. From the earlier discussion, we know that P.808 correlation is highly repeatable between runs when averaged across a set of clips per condition, which can be formed by grouping clips enhanced by a particular SE model or based on other criteria like SNR or reverb RT60 times. The PCC computed on the average of ratings per group across different runs is \textgreater 0.9. We also found that PCC computed on the same clips but from two different P.808 runs is only about 0.5 due to the high rating noise per clip.

Hence, for stack ranking different noise suppressors we evaluate by computing the average of ratings across the entire test set for each model. Therefore, we compute SRCC and PCC between averaged human ratings and averaged DNSMOS per model. SRCC gives us the stack ranking accuracy of various SE models.

\subsection{Results}
\label{ssec:evaluation}
\subsubsection{DNSMOS vs other objective speech quality metrics}
\begin{figure}[t]
 \centering
  \subfloat[MOS vs PESQ]{\includegraphics[width=0.24\textwidth]{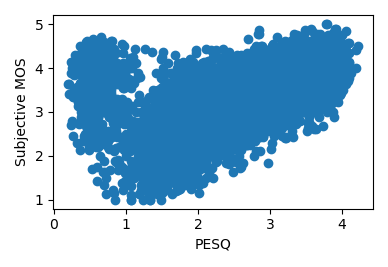}\label{fig:pesq}}
  \hfill
  \subfloat[MOS vs SDR]{\includegraphics[width=0.24\textwidth]{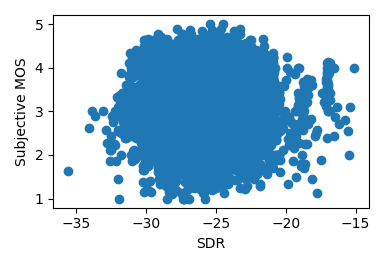}\label{fig:sdr}}
  \hfill
  \subfloat[MOS vs POLQA]{\includegraphics[width=0.24\textwidth]{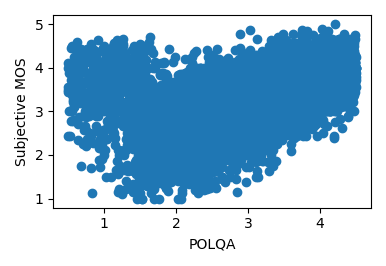}\label{fig:polqa}}
  \hfill
  \subfloat[MOS vs DNSMOS]{\includegraphics[width=0.24\textwidth]{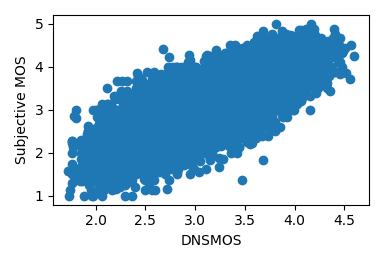}\label{fig:dnsmos}}
  
  \caption{Scatter plots of different objective metrics vs MOS}
\label{fig:scatter}
\end{figure}

Table \ref{tab:table2} shows the PCC and SRCC between human ratings and widely used PESQ, POLQA, SDR and DNSMOS (\(\textbf{\textit{M}}_0\)) computed on the Interspeech DNS challenge blind test results \cite{reddy2020interspeech}. The results show that SDR correlates poorly with MOS. PESQ and POLQA reasonable PCC and SRCC. But the primary DNSMOS model (\(\textbf{\textit{M}}_0\)) correlates significantly better than other objective metrics and is highly reliable in stack ranking the SE models. The scatter plots in Figure. \ref{fig:scatter} shows that DNSMOS aligns with MOS better than SDR, PESQ, and POLQA. Both PESQ and POLQA have a cluster on the left top of the graph indicating that they tend to penalize certain artifacts more than we humans do. 
\begin{figure}[t]
 \centering
    \includegraphics[width=0.7\columnwidth]{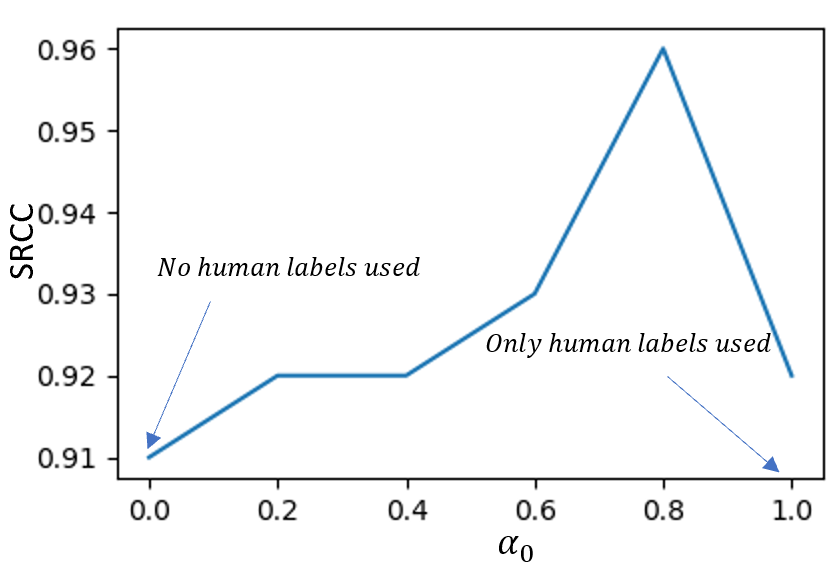}
    \caption{SRCC between MOS and \(\textbf{\textit{M}}_1\) for different \(\alpha_0\) values }  
\label{fig:alpha}
\end{figure}

\begin{table}[!t]
  \begin{center}
    \caption{Per category SRCC between (\(\textbf{\textit{M}}_0\)) and MOS on DNS Challenge 2 results}
    \label{tab:table3}
    \resizebox{\columnwidth}{!}{\begin{tabular}{c|c|c|c|c|c|c}
       \(\textbf{\textit{M}}_i\)(\(\alpha_{0}\),\(\alpha_{1}\),\(\alpha_{2}\)) &\textbf{English} & \textbf{Non-English} &\textbf{Tonal} &\textbf{Emotional} &\textbf{Singing} &\textbf{Overall} \\
      \hline
       \(\textbf{\textit{M}}_0\)(1 ,- ,-) & 0.78 & 0.67 & 0.51 & 0.69 & 0.36 & 0.65\\
      \hline
       \(\textbf{\textit{M}}_1\)(0.8, 0.2, -) & 0.81 & 0.86 & 0.81 & 0.58 & 0.61 & \textbf{0.90}\\
      \hline
      \(\textbf{\textit{M}}_1\)(0.6, 0.4, -) & 0.90 & 0.85 & 0.82 & 0.72 & 0.58 & 0.88\\
      \hline
      \(\textbf{\textit{M}}_1\)(0.4, 0.6, -) & \textbf{0.93} & \textbf{0.92} & \textbf{0.87} & 0.67 & 0.57 & 0.88\\
      \hline
      \(\textbf{\textit{M}}_1\)(0.2, 0.8, -) & 0.90 & 0.85 & 0.82 & 0.72 & 0.58 & 0.88\\
      \hline
      \(\textbf{\textit{M}}_2\)(-, 0.5, 0.5) & 0.90 & 0.86 & 0.81 & 0.75 & 0.54 & \textbf{0.90}\\
      \hline
      \(\textbf{\textit{M}}_2\)(0.6, 0.2, 0.2) & 0.91 & 0.85 & 0.82 & \textbf{0.78} & \textbf{0.72} & 0.89\\
      \hline
      \(\textbf{\textit{M}}_2\)(0.8, 0.1, 0.1) & 0.88 & 0.85 & 0.82 & 0.68 & 0.69 & 0.88\\
      \hline
    \end{tabular}}
  \end{center}
\end{table}

\subsubsection{Generalizability of DNSMOS}
We initially analyzed the effect of \(\alpha_{0}\) on \(\textbf{\textit{M}}_1\) to understand the weightage given to the teacher model \(\textbf{\textit{M}}_0\) vs human ratings. Figure. \ref{fig:alpha} shows the SRCC between MOS and \(\textbf{\textit{M}}_1\) for different values of \(\alpha_{0}\) computed on one of the P.808 runs with 12 closely stack ranked models. We see in Figure. \ref{fig:alpha} that the SRCC is higher for \(\alpha_{0}=0.8\). 

The noisy and the processed clips comprised of a similar distribution as that of the training set. In order to test the generalization capability of training multiple stages of self-teaching models, we evaluated on the ICASSP 2021 DNS Challenge 2 final submissions \cite{reddy2020icassp}, which comprised of the blind test set processed by 18 different noise suppressors. The blind set included emotional speech, singing, utterances in English and non-English languages including a few tonal languages. The DNSMOS models were trained on the data which was predominantly in English with very little or no emotional content such as laughter, crying, yelling, anger, etc. The training set did not include singing and other non-English languages as well. Table \ref{tab:table3} shows the SRCC between the DNSMOS models at various stages and P.808 MOS for each category in the blind set. The SRCC of the primary model \(\textbf{\textit{M}}_0\) is quite low as it does not generalize well for categories other than English. \(\textbf{\textit{M}}_1\) significantly improved SRCC from 0.65 to 0.9 for \(\alpha_{0}=0.8\). The optimal values of \(\alpha_{0}\) were different for each category in table \ref{tab:table3}. We also experimented with \(\textbf{\textit{M}}_2\) for a couple of settings of (\(\alpha_{0}\),\(\alpha_{1}\),\(\alpha_{2}\)). For \(\textbf{\textit{M}}_2\)(-, 0.5, 0.5) with \(\textbf{\textit{M}}_1\)(0.2, 0.8, -) for stage 2 gave the best results. The other settings did not show much improvement over \(\textbf{\textit{M}}_1\).   

\section{Conclusion and Future work}
DNSMOS is a robust speech quality metric designed to stack rank noise suppressors with great accuracy. The multi-stage self-teaching approach significantly improved accuracy. In the future, we would like to deepen our theoretical understanding of how the choice of \(\alpha_{i}\) influences the accuracy of \(\textbf{\textit{M}}_i\) for different datasets. This will help in training DNSMOS on other impairment types such as network distortions, codec artifacts, and reverberation without the need for large scale data with human labels.


\bibliographystyle{IEEEbib}
\bibliography{strings,refs}

\end{document}